\documentclass[twocolumn, pra, showpacs,superscriptaddress]{revtex4}
\usepackage{amssymb}
\usepackage{graphicx}
\usepackage{dcolumn}
\usepackage{bm}
\usepackage{amsmath}

\begin{document}

\title{Ground-state Competition of Two-Component Bosons in Optical Lattice
near a Feshbach Resonance}
\author{Liping Guo}
\affiliation{Department of Physics and Institute of Theoretical Physics, Shanxi
University, Taiyuan 030006, P. R. China}
\affiliation{Institute of Physics, Chinese Academy of Sciences, Beijing 100080, P. R.
China}
\author{Yunbo Zhang}
\affiliation{Department of Physics and Institute of Theoretical Physics, Shanxi
University, Taiyuan 030006, P. R. China}
\affiliation{Laboratory of Optics and Spectroscopy, Department of Physics, University of
Turku, 20014 Turku, Finland}
\author{Shu Chen}
\email{schen@aphy.iphy.ac.cn}
\affiliation{Institute of Physics, Chinese Academy of Sciences, Beijing 100080, P. R.
China}

\begin{abstract}
We investigate the ground state properties of an equal mixture of
two species of bosons in its Mott-insulator phase at a filling
factor two per site. We identify one type of spin triplet-singlet
transition through the competition of ground state. When the
on-site interaction is weak ($U<U_c$) the two particles prefer to
stay in the lowest band and with weak tunnelling between
neighboring sites the system is mapped into an effective spin-1
ferromagnetic exchange Hamiltonian. When the interaction is tuned
by a Feshbach resonance to be large enough ($U>U_c$), higher band
will be populated. Due to the orbital coupling term $S^+S^-$ in
the Hamiltonian, the two atoms in different orbits on a site would
form an on-site singlet. For a non-$SU(2)$-symmetric model,
easy-axis or easy-plane ferromagnetic spin exchange models may be
realized corresponding to phase separation or counter-flow
superfluidity, respectively.
\end{abstract}

 \startpage{1}
\endpage{7}
\maketitle

The study of quantum phase transition in optical lattices has made
great progress both theoretically and experimentally \cite{Jaksch,M.
Greiner} and becomes one of focusing issues of current interest in
the exploration of rich physics in ultracold atomic systems. Jaksch
\textit{et. al.} predicted that the dynamics of a single-component
Bose gas loaded into the lowest band
of an optical lattice is well described by the Bose-Hubbard model \cite%
{Jaksch} and Greiner \textit{et. al.} experimentally confirmed
that the phase transition from superfluid phase to Mott-insulator
could be realized by suppressing tunnelling between neighboring
sites \cite{M. Greiner}. For single-component bosons without
internal degrees of freedom, the superfluid-insulator transition
in a periodic lattice has been extensively studied by various
methods \cite{Jaksch,Stoof,Fisher}. When the spinless bosons are
in the Mott phase, the on-site fluctuation of particle numbers is
suppressed \cite{Fisher}. Many studies have shown that
multi-component bosonic or fermionic gases in optical lattices
exhibit much richer phase diagrams
\cite{conterflow,Duan,commensurate,Fermi,Ehud Altman,Wu}. An
intriguing feature of the multi-component Bose systems is the
structure of their internal ``spin'' degree of freedom. The recent
formation of bound repulsive atom pairs in an optical lattice even
exemplifies stable states without any analogue in traditional
condensed matter physics \cite{Winkler}.

So far, a number of schemes have been proposed to derive an effective
Hamiltonian to describe the spin-related dynamics for the multi-component
system in the Mott-insulator phase\cite{Ehud Altman,Duan}. Most of the
schemes ignore the existence of the upper bands and take single-band
approximation, which is reasonable when the on-site interaction is much
smaller than the energy gap between the first band and the second one. The
situation may change dramatically if the scattering strength of the atoms is
greatly enhanced by the Feshbach resonance so that the on-site interaction
exceeds the band gap. Recently, K\"{o}hl \textit{et. al.} have studied the
fermionic mixture of two hyperfine states of $^{40}$K in a three-dimensional
optical lattice and accessed the strongly interacting regime via a Feshbach
resonance, in which coupling between the lowest energy bands was dynamically
generated \cite{Koehl}. Theoretically, Diener and T.-L. Ho analyzed that a
band insulator may evolve into the state with more bands occupied near the
Feshbach resonance \cite{Diener}. Very recently, A. F. Ho studied the phase
transition from band insulator to Mott insulator for a fermionic system in
optical lattices at a filling of two fermions per site under the two-band
approximation \cite{A. F. Ho}. In that work, the Hund-like orbital coupling
term is shown to play a special role in the strongly interacting regime and
favors spin alignment between different orbits.

It is thus physically nontrivial to go beyond the single-band
approximation. Motivated by the recent progress on the research of
the atomic gas in optical lattice near a Feshbach resonance, in
this paper we study the equal-mixing two-component bosons in
optical lattice with a filling of two bosons per site, focusing on
the Mott-insulating regime and the spin-related phase transition
due to the Feshbach resonance. As in the fermionic case, on each
site there are many orbits and higher orbits may be occupied when
the system is near the Feshbach resonance. Without loss of
generality and for the purpose of simplicity, we take into account
only two bands in the following text, which can be fulfilled by
enforcing the on-site interacting energy smaller than the energy
level spacing between the third and the first orbital. We will
show that in the strongly interacting regime the induced
inter-band coupling prefers the two atoms in different orbits on a
site to form an on-site singlet, which is quite different from the
Hund-like orbital coupling in the fermionic systems \cite{A. F.
Ho}. For simplicity, we consider only a one-dimensional (1D)
system which can be achieved by tuning the laser amplitudes
$V_{0x}\ll V_{0y}, V_{0z}$ to produce a set of uncoupled 1D tubes
\cite{H. Moritz1,H. Moritz2}. In each tube, the system is
effectively described by a 1D optical lattice because the
transverse motion is completely frozen.

We start with the microscopic Hamiltonian of the two-component bosonic
system in a 1D optical lattice
\begin{align}
H& =\sum_{\sigma }\int_{0}^{L}dx\frac{\hbar ^{2}}{2m}\left( \partial
_{x}\psi _{\sigma }^{\dagger }\left( x\right) \partial _{x}\psi _{\sigma
}\left( x\right) \right)  \notag \\
& +\sum_{\sigma }\int_{0}^{L}dx\left( V_{0x}\sin ^{2}kx-\mu _{\sigma
}\right) \psi _{\sigma }^{\dagger }\left( x\right) \psi _{\sigma }\left(
x\right)  \notag \\
& +\sum_{\sigma ,\sigma ^{\prime }}\frac{c}{4}\int_{0}^{L}dx\psi _{\sigma
}^{\dagger }\left( x\right) \psi _{\sigma ^{\prime }}^{\dagger }\left(
x\right) \psi _{\sigma ^{\prime }}\left( x\right) \psi _{\sigma }\left(
x\right) ,  \label{1'}
\end{align}%
where the \textquotedblleft spin\textquotedblright\ indices $\sigma
=\uparrow ,\downarrow $ indicate the two species of atoms or, equivalently,
atoms with two internal states and $\mu _{\sigma }$ is the chemical
potential. For the equal-mixing bosons, we have $\mu _{\uparrow }=\mu
_{\downarrow }=\mu $ corresponding to $N_{\uparrow }=N_{\downarrow }$, where
$N_{\sigma }$ is the total atom number of each specie. The optical lattice
potential has the form of $V_{0x}\sin ^{2}kx$ with wave vectors $k={2\pi }/{%
\ \lambda }$ and $\lambda $ the wavelength of the laser light. The parameter
$c>0$ describes the repulsive interaction of the atoms and the interaction
strengthes of intra-species and of inter-species are taken to be the same.
Since we are interested in the regime where the interaction energy is tuned
so that at most two Bloch bands are populated, it is sufficient to expand
the operator $\psi _{\sigma }\left( x\right) $ in the lowest two Wannier
functions
\begin{equation}
\psi _{\sigma }\left( x\right) =\sum_{i,\alpha =1,2}\omega _{i\alpha }\left(
x\right) c_{i\sigma \alpha },  \label{2'}
\end{equation}%
where the operator $c_{i\sigma \alpha }$ annihilates an atom with spin $%
\sigma $ in the band $\alpha $ at lattice site $i$. In a deep lattice the
Wannier functions $\omega _{i\alpha }\left( x\right) $ can be approximated
by the local harmonic oscillator orbits in the ground state and the first
excited state
\begin{align}
\omega _{i1}\left( x\right) & =\frac{1}{\left( \pi a_{0}^{2}\right) ^{1/4}}%
\exp \frac{-\left( x-x_{i}\right) ^{2}}{2a_{0}^{2}},  \label{3} \\
\omega _{i2}\left( x\right) & =\frac{\left( -1\right) ^{i}}{\left( \pi
a_{0}^{2}\right) ^{1/4}}\frac{\sqrt{2}\left( x-x_{i}\right) }{a_{0}}\exp
\frac{-\left( x-x_{i}\right) ^{2}}{2a_{0}^{2}},  \label{4}
\end{align}%
where $a_{0}=\sqrt{\hbar /m\omega _{T}}$ is the ground state size of the
local harmonic oscillator. Here $\omega _{T}=\sqrt{4V_{0x}E_{R}}/\hbar $ and
$E_{R}=\hbar ^{2}k^{2}/2m$ is the recoil energy.

The second quantized Hamiltonian thus consists of three parts
\begin{equation}
H=H_{t}+H_{intra}+H_{inter}.  \label{123}
\end{equation}%
The hopping term $H_{t}$ describes tunnelling of atoms from one
site to another, which is typically assumed to occur between the
nearest neighboring sites
\begin{equation}
H_{t}=-\sum_{i,\sigma ,\alpha ,\beta }t_{\alpha \beta }c_{i+1\sigma \alpha
}^{\dagger }c_{i\sigma \beta }+H.c.,  \label{1}
\end{equation}%
and the hopping energy is
\begin{equation}
t_{\alpha \beta }=-\frac{\hbar ^{2}}{2m}\int_{0}^{L}dx\partial _{x}\omega
_{i+1\alpha }\left( x\right) \partial _{x}\omega _{i\beta }\left( x\right) .
\label{3'}
\end{equation}%
$H_{intra}$ is the contact-type interaction Hamiltonian in the same energy
band
\begin{align}
H_{intra}& =-\sum_{i,\sigma ,\alpha }\mu _{\alpha }c_{i\sigma \alpha
}^{\dagger }c_{i\sigma \alpha }+\sum_{i,\alpha }U_{\alpha \alpha
}n_{i\uparrow \alpha }n_{i\downarrow \alpha }  \notag \\
& +\frac{1}{2}\sum_{i,\sigma ,\alpha }U_{\alpha \alpha }n_{i\sigma \alpha
}\left( n_{i\sigma \alpha }-1\right) ,  \label{2}
\end{align}%
where the chemical potentials for each band
\begin{equation}
\mu _{\alpha }=-\int_{0}^{L}dx\left(-\frac{\hbar^2}{2m} \partial
_{x}^2 + V_{0x}\sin ^{2}kx-\mu \right) \omega _{i\alpha
}^{2}\left( x\right) \label{4'}
\end{equation}%
are distinguished by a difference $\triangle =\mu _{1}-\mu _{2}$. This
difference is roughly the band gap between the two bands for deep lattice.
On the other hand, the on-site interaction Hamiltonian between the two bands
is denoted as $H_{inter}$
\begin{align}
H_{inter}& =\sum_{i,\alpha \neq \beta }U_{\alpha \beta }\left( n_{i\uparrow
\alpha }n_{i\downarrow \beta }+S_{i\alpha }^{+}S_{i\beta
}^{-}+\bigtriangleup _{i\alpha }^{\dagger }\bigtriangleup _{i\beta }\right)
\notag \\
& +\frac{1}{2}\sum_{\substack{ i,\sigma ,\alpha \neq \beta }}U_{\alpha \beta
}\left( n_{i\sigma \alpha }n_{i\sigma \beta }+\bigtriangleup _{i\sigma
\alpha }^{\prime \dagger }\bigtriangleup _{i\sigma \beta }^{\prime }\right) .
\label{5}
\end{align}%
where $\bigtriangleup _{i\beta }=c_{i\downarrow \beta }c_{i\uparrow \beta }$%
, $\bigtriangleup _{i\sigma \beta }^{\prime }=c_{i\sigma \beta }c_{i\sigma
\beta }$ and $S_{i\alpha }^{+}=c_{i\uparrow \alpha }^{\dagger
}c_{i\downarrow \alpha }$ $\left( S_{i\alpha }^{-}=\left( S_{i\alpha
}^{+}\right) ^{\dagger }\right) $ is a pseudo-spin operator. The repulsive
interaction (positive scattering length) between two atoms sharing a lattice
site in the same band or between the two bands gives rise to an interaction
energy
\begin{equation}
U_{\alpha \alpha }=\frac{c}{2}\int_{0}^{L}dx\omega _{i\alpha }^{4}\left(
x\right) ,  \label{5'}
\end{equation}%
or
\begin{equation}
U_{12}=\frac{c}{2}\int_{0}^{L}dx\omega _{i1}^{2}\left( x\right) \omega
_{i2}^{2}\left( x\right) =U_{21},  \label{6'}
\end{equation}%
which is just the additional energy that one needs to put two atoms on one
site, in the same band or in different bands. The term of $S_{i\alpha
}^{+}S_{i\beta }^{-}$ describes the orbital coupling between the upper and
lower bands/orbits. A striking feature here is that we have got an
interaction term with opposite sign compared to the Fermionic case \cite{A.
F. Ho}, for which the Hund-like orbital coupling term favors the
\textquotedblleft spin\textquotedblright\ of the two fermions at each site
residing in different bands aligning paralleled. The orbital coupling thus
determines the ground state in a different way for the bosonic case. The
spins tend to align anti-paralleled in different bands when the interaction
exceeds the energy gap far away as illustrated later in Figure 2. The terms
of $\bigtriangleup _{i\alpha }^{\dagger }\bigtriangleup _{i\beta }$ and $%
\bigtriangleup _{i\sigma \alpha }^{\prime \dagger }\bigtriangleup _{i\sigma
\beta }^{\prime }$ describe the interaction of atomic pair in different
bands.

Substituting the approximate Wannier functions Eqs. (\ref{3}) and (\ref{4})
into (\ref{3'}), (\ref{5'}) and (\ref{6'}) we easily obtain the parameters $%
U_{11}=U, U_{22}=0.75U, U_{12}=0.5U$ where $U=c/4\sqrt{2\pi }a_{0}$. Unlike
the long-range Coulomb interactions for electrons in solids, here the
orbital coupling term is of the same order of magnitude as the on-site
repulsion term. Owing to the approximation of local harmonic oscillator
orbits on Wannier functions, the integral of the hopping matrix element
between different bands is nonzero and in fact they satisfy the relations $%
\left\vert t_{11}\right\vert <\left\vert t_{12}\right\vert <\left\vert
t_{22}\right\vert $. In optical lattice, both the hopping term $t_{\alpha
\beta }$ and the on-site interaction $U$ depend on the amplitude $V_{0}$ of
the laser field. In this work, we will focus on the Mott phase with a larger
ratio of $U/t_{\alpha \beta }$ and study the ground-state phase transition
due to the change of the on-site interaction. In principle, via the Feshbach
resonance, one could tune the strength of interaction so that $U<\triangle $
or $\triangle <U<2\triangle $. In the former case two bosons occupy the
lowest Bloch band while in the latter case one of the atoms in the lowest
band would be forced into the higher excited band.

In the strong coupling limit with $t_{\alpha \beta }\ll U_{\alpha \beta
},\Delta $, it is instructive to first consider the on-site local
Hamiltonian with $t_{\alpha \beta }=0$ and then treat the hopping term (\ref%
{1}) as perturbation. It is easy to diagonalize the local on-site
Hamiltonian (both intra- and inter-band parts (\ref{2}) and (\ref{5}))with
two bosons per site. The local spectrum are given by:
\begin{eqnarray*}
\varepsilon _{1} &=&-2\mu _{1}+\triangle +\frac{7U}{8}-\sqrt{\left(
\triangle -\frac{U}{8}\right) ^{2}+\left( \frac{U}{2}\right) ^{2}}, \\
\varepsilon _{2} &=&-2\mu _{1}+\triangle , \\
\varepsilon _{3} &=&-2\mu _{1}+\triangle +\frac{U}{2}, \\
\varepsilon _{4} &=&-2\mu _{1}+\triangle +U, \\
\varepsilon _{5} &=&-2\mu _{1}+\triangle +\frac{7U}{8}+\sqrt{\left(
\triangle -\frac{U}{8}\right) ^{2}+\left( \frac{U}{2}\right) ^{2}}.
\end{eqnarray*}

\begin{figure}[tbp]
\includegraphics[width=3.5in]{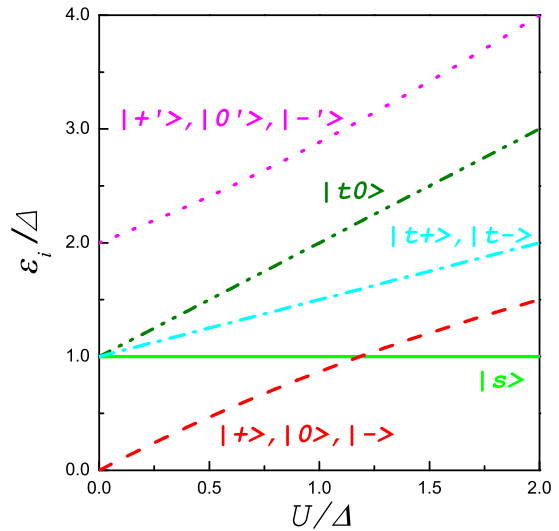}
\caption{(Color online) The on-site energies $\protect\varepsilon_{i}$
versus $U/\triangle$ where we have made a total energy shift of $-2\protect%
\mu_{1}$.}
\label{fig1}
\end{figure}

Among the ten eigenstates, those corresponding to eigenenergy $\varepsilon
_{1}$ are three-fold degenerate and are given by
\begin{eqnarray*}
\left\vert +\right\rangle &=&-e\left\vert \uparrow \uparrow ,0\right\rangle
+f\left\vert 0,\uparrow \uparrow \right\rangle , \\
\left\vert 0\right\rangle &=&-e\left\vert \uparrow \downarrow
,0\right\rangle +f\left\vert 0,\downarrow \uparrow \right\rangle . \\
\left\vert -\right\rangle &=&-e\left\vert \downarrow \downarrow
,0\right\rangle +f\left\vert 0,\downarrow \downarrow \right\rangle;
\end{eqnarray*}
The state corresponding to $\varepsilon _{2}$ is a local singlet forming by
the atoms in the upper and lower orbits
\begin{equation*}
\text{ }\left\vert s\right\rangle =\frac{1}{\sqrt{2}}\left( \left\vert
\uparrow ,\downarrow \right\rangle -\left\vert \downarrow ,\uparrow
\right\rangle \right);
\end{equation*}%
The states corresponding to $\varepsilon _{3}$ are
\begin{eqnarray*}
\left\vert t+\right\rangle &=&\left\vert \uparrow ,\uparrow \right\rangle ,
\\
\left\vert t-\right\rangle &=&\left\vert \downarrow ,\downarrow
\right\rangle ,
\end{eqnarray*}%
and state corresponding to $\varepsilon _{4}$ is
\begin{equation*}
\left\vert t0\right\rangle =\frac{1}{\sqrt{2}}\left( \left\vert \uparrow
,\downarrow \right\rangle +\left\vert \downarrow ,\uparrow \right\rangle
\right);
\end{equation*}%
Finally the states corresponding to $\varepsilon _{5}$ are again three-fold
degenerate
\begin{eqnarray*}
\left\vert +^{\prime }\right\rangle &=&f\left\vert \uparrow
\uparrow
,0\right\rangle +e\left\vert 0,\uparrow \uparrow \right\rangle , \\
\left\vert 0^{\prime }\right\rangle &=&f\left\vert \uparrow
\downarrow
,0\right\rangle +e\left\vert 0,\downarrow \uparrow \right\rangle , \\
\left\vert -^{\prime }\right\rangle &=&f\left\vert \downarrow
\downarrow ,0\right\rangle +e\left\vert 0,\downarrow \downarrow
\right\rangle .
\end{eqnarray*}
Here we have used the notation for the representation of eigenstates that
the left of comma in the right bar is for band 1 and the right of comma is
for band 2. For example, $\left\vert \uparrow \uparrow ,0\right\rangle =%
\frac{1}{\sqrt{2}}\left( c_{\uparrow 1}^{\dagger }\right) ^{2}\left\vert
0\right\rangle $ represents two atoms with spin of $\uparrow $ in the lower
orbit and $\left\vert \uparrow ,\downarrow \right\rangle =c_{\uparrow
1}^{\dagger }c_{\downarrow 2}^{\dagger }\left\vert 0\right\rangle $
represents an atom with spin of $\uparrow $ in the lower orbit and an atom
with spin of $\downarrow $ in the upper orbit. The coefficients
\begin{eqnarray*}
e &=&\frac{1}{\sqrt{2}}\sqrt{1+\frac{1}{\sqrt{1+\left( \frac{1}{2\frac{%
\triangle }{U}-\frac{1}{4}}\right) ^{2}}}}, \\
f &=&\frac{1}{\sqrt{2}}\sqrt{1-\frac{1}{\sqrt{1+\left( \frac{1}{2\frac{%
\triangle }{U}-\frac{1}{4}}\right) ^{2}}}},
\end{eqnarray*}%
fulfill $e^{2}+f^{2}=1$ and $e^{2}$ and $f^{2}$ describe the probability two
atoms simultaneously stay at the lowest band and the upper band
respectively. We have $0<f^{2}\lesssim 0.0659$ for $0<U<\triangle $. When $%
U/\triangle \rightarrow 0$, $f^{2}\rightarrow 0$ and thus the system goes
back to the single band model. To give concrete examples, we note $%
f^{2}=0.0006$ for ${U}/{\triangle }=0.1$ and $f^{2}=0.0169$ for ${U}/{%
\triangle }=0.5$. Hence in the weakly interacting regime the two atoms
mainly stay in the lowest band.

In Figure 1, we display the five eigenenergies as a function of $U/\Delta $.
To get the phase diagram for the ground state, it is sufficient to identify
the lowest two levels while $\varepsilon _{3}$, $\varepsilon _{4}$ and $%
\varepsilon _{5}$ always correspond to the higher bands. The competition of
the lowest two levels gives rise to completely different ground state
structure of the system and the transition point ${U_{c}}/{\ \triangle }%
\simeq 1.19$ is approximately determined by the energy level crossing of $%
\varepsilon _{1}$ and $\varepsilon _{2}$. For $U>U_{c}$, the local ground
state on each site is a singlet state $\left\vert s\right\rangle $ with the
spins of the two bosons aligned anti-paralleled. For $U<U_{c}$, the local
ground state is one of the spin triplet $\left\vert +\right\rangle $, $%
\left\vert 0\right\rangle $ and $\left\vert -\right\rangle $. It is
worthwhile to indicate that although the total spin fulfils $\left\langle
S_{z}^{total}\right\rangle =0$ as a result of $N_{\uparrow }=N_{\downarrow }$%
, at each site the two species of bosons are not necessarily equal-mixing.
At the first sight, this seems to imply that, in the limit of $t_{\alpha
\beta }=0$ and $U<U_{c}$, the ground state of the whole system is highly
degenerate and the spins of atoms at each site align arbitrarily because the
local ground state on an isolated site can be either of the three states as
long as the total spin of the system is zero. Actually this is not true when
the hopping processes between the neighboring sites are considered.

Now we switch on the hopping term between the nearest-neighbor sites. For
the system with a filling factor two, the state with two atoms at each site
has lowest on-site energy. The process of an atom hopping to its neighboring
site would change the on-site population, however such a hopping process is
greatly suppressed because placing three atoms at a site extremely costs
energy. Nevertheless, the virtual process of hopping to an intermediate
state and then hopping back gives a second order correction to the ground
state energy and lowers the ground state energy. The virtual hopping process
does not change the total on-site populations but can exchange two different
atoms on neighboring sites. These virtual exchanging processes can be
described by an effective Hamiltonian acting on the ground states which is
obtained in a second-order perturbation theory as
\begin{equation}
H_{eff}= \sum_{i,m} \frac{\left\langle \mu_{i,i+1} \right \vert H_{t}
\left\vert m\right\rangle \left\langle m \right \vert H_{t} \left\vert
\upsilon_{i,i+1} \right\rangle} {E_{0}-E_{m}} \left\vert \mu_{i,i+1}
\right\rangle \left\langle \upsilon_{i,i+1} \right \vert  \label{14}
\end{equation}
where $\{\left\vert \mu_{i,i+1} \right\rangle,\left\vert \upsilon_{i,i+1}
\right\rangle = \left\vert g \right\rangle_i \otimes \left\vert
g\right\rangle_{i+1} \}$ are ground states with $N_i = N_{i+1} =2$ and with
ground state energy $E_{0}$. As $U<U_c$, the three-fold degenerate ground
states at an isolated site form a triplet, i.e., $\left\vert g
\right\rangle_i =\left\vert +\right\rangle_i,\left\vert 0\right\rangle_i,$
or $\left\vert -\right\rangle_i$, and therefore $\left\vert \mu_{i,i+1}
\right\rangle$ is nine-fold degenerate with $E_0 = 2\varepsilon_1$. On the
opposite regime of $U>U_c$, $\left\vert \mu_{i,i+1} \right\rangle=\left\vert
s \right\rangle_i \otimes \left\vert s \right\rangle_{i+1}$ with $E_0 =
2\varepsilon_2$. The intermediate states are the product of states on the
two neighboring sites with three and one bosons respectively and with
excitation energies $E_{m}$.

\begin{figure}[ptb]
\includegraphics[width=3in]{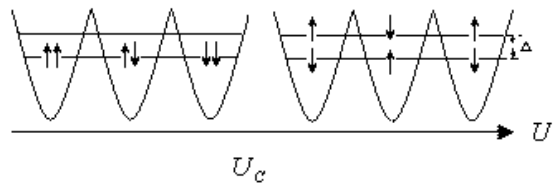}
\caption{Schematic picture for competing ground states in optical lattice. A
phase transition from spin exchange to bosonic singlet occurs at $U=U_c$}
\label{fig2}
\end{figure}

The second order perturbation calculation of the hopping terms
enables us to identify one type of spin related quantum phase
transition induced by the Feshbach resonance. On the one side of
the transition point, that is in the weakly interacting regime
($U<U_{c}$), the effective Hamiltonian can be further simplified
and represented as an effective isotropic Heisenberg model in
terms of spin-1 operators. After straightforward but tedious
algebra we get the effective Hamiltonian
\begin{equation}
H_{eff}=-\lambda \sum_{\left\langle i,j\right\rangle }\mathbf{S}_{i}\cdot
\mathbf{S}_{j}  \label{ss}
\end{equation}%
where $S^{\alpha }$ is a spin-1 operator in $\alpha \left( \alpha
=x,y,z\right) $ orientation and the spin exchange coefficient is
\begin{equation}
\lambda =e^{4}\frac{2\left( t_{11}\right) ^{2}}{U}+2e^{2}f^{2}\left( \frac{%
\left( t_{12}\right) ^{2}}{\triangle +U}+\frac{\left( t_{12}\right) ^{2}}{%
3\triangle +\frac{1}{4}U}\right) .
\end{equation}%
where terms $f^{4}$ is neglected due to its smallness. Eq.
(\ref{ss}) is nothing but the Hamiltonian of an isotropic $S=1$
ferromagnetic quantum Heisenberg spin system. In the limiting case
$\triangle \gg U$, that is, when the upper band lies much higher
than the lower band, we find
\begin{equation}
\lambda =2\frac{\left( t_{11}\right) ^{2}}{U}.
\end{equation}%
In this limit, the probability of two atoms occupying the lowest
band $e^{2}$ approaches unity. We then recover the result in the
single band approximation \cite{conterflow,Ehud Altman}. The
isotropic ferromagnetic model (\ref{ss}) has ($2S^{total}+1$)-fold
degeneracy with $S^{total}=N/2$ ($N/2=N_{\uparrow}=N_{\downarrow}$).
The ground state corresponds to the state with $S^{total}_z =0$, in
which case no spatial broken symmetry occurs. In the bosonic
langauge, this means that the system phase does not separate in the
ground state for the $SU(2)$-symmetric model.

On the other side of the transition point, when $U>U_{c}$, the ground state
at the isolated site is a singlet. In this case, the virtual hopping process
does not induce redistribution of on-site spins and the global ground state
is the product of on-site singlets. We straightforwardly obtain the
correction to ground state energy per site of the optical lattice by
calculating the virtual hopping process to second order perturbation
\begin{equation}
\epsilon =\varepsilon _{2}+\delta \epsilon
\end{equation}%
with
\begin{equation}
\delta \epsilon =-\frac{3}{2}\left( \frac{(t_{11})^{2}}{2U}+\frac{%
(t_{22})^{2}}{\frac{7}{4}U}+\frac{(t_{12})^{2}}{\triangle +\frac{7}{4}U}+%
\frac{(t_{12})^{2}}{2U-\triangle }\right) .
\end{equation}%
Obviously this correction is negative and the hopping process
always lowers the ground state energy.

Figure 2 depicts the phase diagram of two-boson in two-band optical
lattice model. For the interaction $U<U_{c}$, the atoms on a site
form a triplet and the virtual hopping process produces
ferromagnetic exchange between spins on neighboring sites, while in
the strong coupling limit $ U>U_{c}$ the atoms in different bands
prefer to align their spin anti-paralleled and form an on-site
singlet. A phase transition from spin exchange to bosonic singlet
occurs therefore at $U=U_{c}$. We recall that in the fermionic case
the phase diagram exhibits drastically different structure. Fermions
with a filling factor two in two-band optical lattice are shown to
exhibit opposite behavior and there exists a phase transition from
the band insulator to a Mott insulator with interesting dynamics of
a spin-1 Heisenberg anti-ferromagnet \cite{A. F. Ho}.

We notice here a big difference between 1D fermions and bosons.
According to Haldane's conjecture, the ground state of
$SU(2)$-symmetric anti-ferromagnetic spin-1 Hamiltonian is gapped.
Hence, small deviations of the Hamiltonian parameters reducing the
$SU(2)$ symmetry to $U(1)$ will not lead to qualitatively different
results. On the other hand, the ground state for the ferromagnetic
Hamiltonian is ordered and it is crucially important to consider a
generic non-$SU(2)$-symmetric model in order to understand whether
the ground state is an easy-axis or easy-plane ferromagnet.

To do this, we let the tunnelling matrix elements $t_{\alpha \beta
}^{\sigma}$ ($\sigma=\uparrow,\downarrow$) depend not only on band
indices $\alpha,\beta$ but also on the component index $\sigma$.
Furthermore we distinguish the intra-species interaction
$U=U_{\uparrow\uparrow}=U_{\downarrow\downarrow}$ and
inter-species interaction $U'=U_{\uparrow\downarrow}$ to break the
$SU(2)$ symmetry. When the system is in the strongly interacting
regime, deviation of the $SU(2)$ symmetry does not lead to
qualitative change of the ground state properties because the
ground state is composed of on-site singlets. However, in the
weakly interacting regime, when the $SU(2)$ symmetry is broken,
the effective Hamiltonian can be of the easy-axis type or of the
easy-plane type with different kinds of ground states. We note
that, for the general case with $U \neq U'$, the effective
Hamiltonian can not be represented in the form of a simple spin
exchange model. However, if $|U'-U| \ll U,U' $, we can attribute
the difference of the on-site interacting energies to the
zeroth-order Hamiltonian \cite{conterflow} and get an effective
Hamiltonian of XXZ model
\begin{eqnarray}
H &=&-\sum_{\left\langle i,j\right\rangle }\left[ \lambda'
\mathbf{S}_{i}\cdot \mathbf{S}_{j}+\delta\lambda' _{z}
S_{iz}S_{jz} \right]    \nonumber\\
&&+B\sum_{i}S_{iz}+D\sum_{i}\left( S_{iz}\right) ^{2} ,
\label{XXZ}
\end{eqnarray}%
where
\begin{equation*}
\lambda'=2e^{4}\frac{t_{11}^{\downarrow }t_{11}^{\uparrow}}{U}
+2e^{2}f^{2}t_{12}^{\uparrow }t_{12}^{\downarrow }\left(
\frac{1}{3\triangle + \frac{1}{4}U}+\frac{1}{\triangle +U}\right)
,
\end{equation*}
\begin{eqnarray*}
\delta\lambda' _{z} &=&e^{4}\frac{\left( t_{11}^{\uparrow }-
t_{11}^{\downarrow }\right) ^{2}}{U} \\
&&+e^{2}f^{2}\left( t_{12}^{\uparrow} - t_{12}^{\downarrow
}\right) ^{2} \left( \frac{1}{\triangle +U}+\frac{1}{ 3\triangle
+\frac{1}{4}U}\right) ,
\end{eqnarray*}
\begin{eqnarray*}
B &=&-e^{4}\left[ 3\frac{ \left( t_{11}^{\uparrow }\right)
^{2}-\left( t_{11}^{\downarrow}\right) ^{2} }{U}+\frac{\left(
t_{12}^{\uparrow
}\right) ^{2}-\left( t_{12}^{\downarrow }\right) ^{2}}{\triangle }\right]  \\
&&-e^{2}f^{2}\left[ \frac{\left( t_{11}^{\uparrow }\right)
^{2}-\left( t_{11}^{\downarrow }\right) ^{2}}{2\triangle
-\frac{1}{4}U}+\frac{\left( t_{22}^{\uparrow }\right) ^{2}-\left(
t_{22}^{\downarrow }\right) ^{2}}{2\triangle
}\right]  \\
&&-3e^{2}f^{2}\left[ \left( t_{12}^{\uparrow }\right) ^{2}-\left(
t_{12}^{\downarrow }\right) ^{2}\right] \left( \frac{1}{3\triangle
+\frac{1}{4}U }+\frac{1}{\triangle +U}\right) ,
\end{eqnarray*}
and
\begin{eqnarray*}
D &=&\frac{7}{8}\left( U-U^{\prime }\right) -\sqrt{\left( \triangle -\frac{U%
}{8}\right) ^{2}+\left( \frac{U}{2}\right) ^{2}} \\
&&+\sqrt{\left( \triangle -\frac{U^{\prime }}{8}\right) ^{2}+\left( \frac{%
U^{\prime }}{2}\right) ^{2}}.
\end{eqnarray*}%
In the limiting case $\triangle \gg U$, it is easy to show that $
\lambda'={2t_{11}^{\downarrow }t_{11}^{\uparrow }}/{U}$,
$\delta\lambda' _{z} = [ t_{11}^{\uparrow }- t_{11}^{\downarrow }]
^{2}/{U}$,  $ B=-{3[( t_{11\uparrow }) ^{2}-\left( t_{11\downarrow
}\right) ^{2}] }/{U}$ and $ D\approx U-U^{\prime }$ and we recover
the result for the single band approximation \cite{conterflow,Ehud
Altman}. It is obvious that we have always a positive small
anisotropy parameter $\delta\lambda' _{z}$ for $t_{\alpha \beta
}^{\uparrow}\neq t_{\alpha \beta }^{\downarrow}$, which implies the
effective XXZ model describes an easy-axis ferromagnet. Under the
condition of $\delta\lambda' _{z}\gg D$, the ground state of the
spin system is in a phase with spin domains. In the bosonic
language, it corresponds to the situation with phase separation of
the two components. This implies that differentiating the tunneling
terms for different components would induce phase segregation. When
$t_{\alpha \beta}^{\uparrow }=t_{\alpha \beta}^{\downarrow }$, we
have $\delta\lambda' _{z}=0$ and $B=0$ which reduces the model to
(\ref {ss}) except an additional term $D$ (which vanishes naturally
for $SU(2)$-symmetric model because $U=U'$). For a large positive
$D$, however, an easy-plane ground state can be realized. In terms
of the nomenclature in Ref. \cite{conterflow}, the easy-plane
ferromagnet means the counter-flow superfluid. Straightforwardly, a
positive $D$ reduces the $S_z$ component of the spin on each site.
At large enough $D>0$, all spins will be essentially confined to the
state with $\langle S_{iz} \rangle=0$, which implies that large
enough intra-component interaction ($U \gg U'$) leads to two atoms
belonging to distinct species occupying each site. On the other
hand, for small enough $D<0$ ($U \ll U'$), the ground state would
stay in the state with $\langle S_{iz} \rangle=\pm 1$ and the term
of $D$ enhances the phase separation of different components.

Before ending the discussion, we would like to remark the
extension of the present work to the case with higher dimensions.
Unlike the single band model which can be directly extended to the
high-dimensional case, the effective Hamiltonians (\ref{ss}) and
(\ref{XXZ}) are no longer applicable to the high-dimensional
optical lattice models when the higher orbits are populated. For
higher dimensions, the first excited state in a local site is
degenerate and has spatial anisotropy. Correspondingly, the
hopping matrix element acquires spatial anisotropy and new
physical phenomena may arise due to the orbital degeneracy
\cite{Isacsson,Kuklov06,Liu}.

In summary, we have studied the quantum phase transition induced by
effective orbital coupling in optical lattices for an equal-mixing
two-component boson system at a filling factor two per site.  In the
regime with weak on-site interaction, the two atoms stay in the
lowest band and can be described by an effective spin-1
ferromagnetic exchange model. In the regime with strong on-site
interaction, the two atoms prefer to occupy different orbits on a
site and form an on-site singlet due to the effective orbital
coupling. We also considered the generic non-$SU(2)$-symmetry model.
In the weakly interacting regime, the ground state may be described
by an easy-axis ferromagnet corresponding to the case of phase
separation or an easy-plane ferromagnet corresponding to the state
of counter-flow superfluid.

S.C. is supported by NSF of China under Grant No. 10574150 and
program of Chinese Academy of Science. Y.Z. is supported by Shanxi
Province Youth Science Foundation under grant No. 20051001 and by
Academy of Finland under grant number 206108.

\end{document}